\documentclass[fleqn, usenatbib, onecolumn]{mnras}

\usepackage[utf8]{inputenc} 
\usepackage[T1]{fontenc}    
\usepackage{hyperref}       
\usepackage{url}            
\usepackage{booktabs}       
\usepackage{amsfonts}       
\usepackage{nicefrac}       
\usepackage{amsmath}
\usepackage{amssymb}
\usepackage{multirow}
\usepackage{float}
\usepackage{graphicx}
\usepackage{cleveref}
\usepackage{physics}
\usepackage{newtxtext,newtxmath}

\newcommand{\tilxi}{\Tilde{\xi}} 
\newcommand{\be}{\begin{equation}} 
\newcommand{\ee}{\end{equation}} 
\newcommand{\bary}{\begin{eqnarray}} 
\newcommand{\eary}{\end{eqnarray}} 
\Crefname{equation}{Eq.}{Eqs.} 

\title[Polarization From Radially Stratified GRB Outflow]{Polarization From A Radially Stratified Off-Axis GRB Outflow}

\author[Caligula et al.]{A. C. Caligula Do E. S. Pedreira, $^{1}$
N. Fraija,$^{1}$ 
A. Galvan-Gamez,$^{1}$
B. Betancourt Kamenetskaia,$^{2,3}$\newauthor
S. Dichiara,$^{4}$
M.G. Dainotti,$^{5,6,7,8}$
R.~L.~Becerra,$^{9}$
and P. Veres$^{10}$
\\
$^{1}$Instituto de Astronom\'ia, Universidad Nacional Aut\'onoma de M\'exico, Circuito Exterior, C.U., A. Postal 70-264, 04510, CDMX, Mexico\\
$^{2}$TUM Physics Department, Technical University of Munich, James-Franck-Str, 85748 Garching, Germany\\
$^{3}$Max-Planck-Institut für Physik (Werner-Heisenberg-Institut), Föhringer Ring 6, 80805 Munich, Germany\\
$^{4}$Department of Astronomy and Astrophysics, The Pennsylvania State University, 525 Davey Lab, University Park, PA 16802, USA\\
$^{5}$Division of Science, National Astronomical Observatory of Japan, 2-21-1 Osawa, Mitaka, Tokyo 181-8588, Japan \\
$^{6}$The Graduate University for Advanced Studies (SOKENDAI), 2-21-1 Osawa, Mitaka, Tokyo 181-8588, Japan\\
$^{7}$Space Science Institute, 4750 Walnut Street, Boulder, CO 80301, USA\\
$^{8}$SLAC National Accelerator Laboratory, 2575 Sand Hill Road, Menlo Park, CA 94025, USA\\
$^{9}$Instituto de Ciencias Nucleares, Universidad Nacional Aut\'onoma de M\'exico, Apartado Postal 70-264, 04510 M\'exico, CDMX, Mexico\\
$^{10}$Center for Space Plasma and Aeronomic Research (CSPAR), University of Alabama in Huntsville, Huntsville, AL 35899, USA\\
}

\date{Accepted XXX. Received YYY; in original form ZZZ}
\pubyear{2022}

\begin{document}
\label{firstpage}
\pagerange{\pageref{firstpage}--\pageref{lastpage}}
\maketitle

\begin{abstract}
While the dominant radiation mechanism gamma-ray bursts (GRBs) remains a question of debate, synchrotron emission is one of the foremost candidates to describe the multi-wavelength afterglow observations. As such, it is expected that GRBs should present some degree of polarization across their evolution -- presenting a feasible means of probing these bursts’ energetic and angular properties. Although obtaining polarization data is difficult due to the inherent complexities regarding GRB observations, advances are being made, and theoretical modeling of synchrotron polarization is now more relevant than ever. In this manuscript, we present the polarization for a fiduciary model where the synchrotron forward-shock emission evolving in the radiative-adiabatic regime is described by a radially stratified off-axis outflow. This is parameterized with a power-law velocity distribution and decelerated in a constant-density and wind-like external environment. We apply this theoretical polarization model for selected bursts presenting evidence of off-axis afterglow emission, including the nearest orphan GRB candidates observed by the \emph{Neil Gehrels Swift Observatory} and a few Gravitational Wave (GWs) events that could generate electromagnetic emission. In the case of GRB 170817A, we require the available polarimetric upper limits in radio wavelengths to constrain its magnetic field geometry.
\end{abstract}
\begin{keywords}
Physical data and processes: polarization -- (stars:) gamma-ray burst: general
 -- (stars:) gamma-ray burst: individual:...
 -- Physical data and processes:acceleration of particles -- Physical data and processes:magnetic fields
\end{keywords}

\section{Introduction}\label{sec1}

Gamma-ray Bursts (GRBs) are the most luminescent phenomena in the universe. They result from the deaths of massive stars \citep{1993ApJ...405..273W,1998ApJ...494L..45P, 2006ARA&A..44..507W,Cano2017} or the merger of two compact objects, such as neutron stars \citep[NSs;][]{1992ApJ...392L...9D, 1992Natur.357..472U, 1994MNRAS.270..480T, 2011MNRAS.413.2031M} or a NS with a black hole \citep[BH,][]{1992ApJ...395L..83N}. GRBs are evaluated based on the phenomenology seen during their early and late phases and are often characterized by the fireball model \citep{1998ApJ...497L..17S} to distinguish their various sources. The principal and earliest emission, known as the ``prompt emission", is detected from hard X-rays to $\gamma$-rays. This phase can be explained by the interactions of internal shells of material launched forcefully from the central engine at various speeds \citep{1994ApJ...430L..93R, 1994ApJ...427..708P}, photospheric emission from the fireball \citep{2007ApJ...666.1012T,2013ApJ...765..103L,2011ApJ...732...26M} or discharges from a Poynting-flux dominated ejecta \citep{2008A&A...480..305G,2016MNRAS.459.3635B, 2015MNRAS.453.1820K, 2011ApJ...726...90Z}. Later emission, known as ``afterglow", \citep[e.g.,][]{1997Natur.387..783C, 1998ApJ...497L..17S,2002ApJ...568..820G, 1997Natur.386..686V,1998A&A...331L..41P,Gehrels2009ARA&A,Wang2015} is a long-lasting multi-wavelength emission detectable in gamma-rays, X-rays, optical, and radio. It is modeled using synchrotron radiation produced when the external environment decelerates the relativistic outflow, and a significant portion of its energy is transferred. Long GRBs (lGRBs) and short GRBs (sGRBs) are categorized based on their duration:\footnote{For a debate of controversial situations, see \cite{kann2011}.} $T_{90}\leq 2\mathrm{\,s}$ or $T_{90} \ge 2\mathrm{\,s}$,\footnote{$T_{90}$ is the time over which a GRB releases from $5\%$ to $95\%$ of the total measured counts.} respectively \citep{mazets1981catalog, kouveliotou1993identification}. 

Synchrotron radiation is the fundamental emission mechanism in GRB afterglows in a forward-shock (FS) scenario \citep{Kumar,1997ApJ...476..232M}. Nevertheless, synchrotron is contingent on the existence of magnetic fields. The origin and arrangement of these magnetic fields behind the shock remain debatable.
They can originate from the compression of an existing magnetic field within the interstellar medium \citep[ISM;][]{1980MNRAS.193..439L,2021MNRAS.507.5340T} and shock-generated two-stream instabilities \citep{PhysRevLett.2.83, Medvedev}. The magnetic field generated by these plasma instabilities is random in orientation but mostly confined to the plane of the shock \citep{2020MNRAS.491.5815G}. Modeling the source and arrangement of those fields and other physical properties of GRBs presents a challenging task. This has necessitated the development of other methods for investigating these complicated systems. Among these techniques is linear polarization.

Linear polarization has been measured, up to a few percent, from the afterglow of several GRBs. Some examples include GRB 191221B \citep[$\Pi=1.2\%$;][]{Buckley} at the late afterglow, GRB 190114C \citep[$\Pi=0.8\pm0.13\%$;][]{Laskar_2019} on the radio band, and the upper limits determinations of GRB 991216 \citep[yielding $\Pi < 7\%$;][]{2005ApJ...625..263G} and GRB 170817A \citep[yielding $\Pi < 12\%$, on the 2.8 GHz radio band][]{2018ApJ...861L..10C}. 
Since the degree of polarization relies on the configuration of the magnetic field, analyzing the degree of polarization permits us to investigate these configurations and, therefore, their origins. Previous works, including \cite{2003ApJ...594L..83G, Gill-1, 2004MNRAS.350.1288R, Lyutikov, Nakar, 2021MNRAS.507.5340T, 2020ApJ...892..131S}, have already investigated the practicality of utilizing polarization models to acquire source-related information.
Due to the unfortunate short number of orbital polarimeters and the normal difficulty of seeing these extreme events, collecting polarization data has been one of the most significant impediments. Despite this, progress has been made in the field as a result of initiatives like the POLAR project \citep{POLAR}, and it is anticipated that we will have abundant data to test various models in the coming years.

This study expands the analytical synchrotron afterglow scenario of the off-axis homogeneous jet in a stratified environment, which was required to characterize the multi-wavelength data of GRB 170817A \citep{2020ApJ...896...25F} and a sample of GRBs exhibiting off-axis emission.\footnote{We use the values of the cosmological constants $H_0=69.6\,{\rm km\,s^{-1}\,Mpc^{-1}}$, $\Omega_{\rm M}=0.286$ and $\Omega_\Lambda=0.714$ \citep{2016A&A...594A..13P}, which correspond to a spatially flat universe $\Lambda$CDM model.} The phenomenological model is extended from adiabatic to radiative regime, including the self-absorption synchrotron phase and the dimensionless factor, which provides information on the equal arrival time surface (EATS). We show the temporal development of polarization from the synchrotron afterglow stratification model and compute the expected polarization for GRB 080503 \citep{2009ApJ...696.1871P,2015ApJ...807..163G}, GRB 140903A \citep{2016ApJ...827..102T, 2017ApJ...835...73Z}, GRB 150101B \citep{2018NatCo...9.4089T}, GRB 160821B \citep{2019MNRAS.489.2104T} and GRB 170817A \citep{2017Sci...358.1559K, 2017MNRAS.472.4953L, mooley, 2018ApJ...867...95H, 2019ApJ...884...71F}.
For GRB 170817A in particular, we employ the available polarimetric upper limits from \cite{2018ApJ...861L..10C}.
Furthermore, taking into account the multi-wavelength upper limits of the closest Swift-detected bursts and the Gravitational Wave (GW) events that potentially produce electromagnetic emission,\footnote{These events were associated to at least one NS by Advanced Laser Interferometer Gravitational-Wave Observatory (LIGO) and Advanced VIRGO detectors \citep{2021PhRvX..11b1053A, 2021arXiv211103606T}.} we create a polarization curve in order to constrain some of the parameters of our off-axis jet model.
Keeping this in mind, the following is the structure of the paper: In Section \ref{sec2}, we briefly describe the off-axis jet synchrotron model derived in \cite{2020ApJ...896...25F} with the extension. In Section \ref{sec3}, we introduce the polarization model used in this paper. In Section \ref{sec4}, we compute the assumed polarization and give the outcomes for a sample of off-axis afterglow-emitting bursts. In Section \ref{sec5} and Section \ref{sec6}, we give analogous analyses for the closest Swift-detected bursts and the GW events that could have emitted an electromagnetic signature, respectively. Finally, in Section \ref{sec7}, we present the conclusion and provide closing thoughts.

\section{Synchrotron Forward-shock model from a radially stratified off-axis jet}\label{sec2}

The multi-wavelength afterglow observations of GRB 170817A are consistent with the synchrotron FS scenario in the fully adiabatic regime from a radially stratified off-axis outflow decelerated in a homogeneous medium \citep{2019ApJ...871..123F}. \cite{2020ApJ...896...25F} extended the synchrotron FS approach to a stratified environment based on the immediate vicinity of a binary NS merger proposed to explain the gamma-ray flux in GRB 150101B. Additionally, \cite{2020ApJ...896...25F} successfully explained the multi-wavelength afterglow observations in GRB 080503, GRB 140903A and GRB 160821B using the synchrotron off-axis model.

In order to present a polarization model and perform a fully time-evolving analysis, we extend the synchrotron scenario described in \cite{2019ApJ...871..123F, 2020ApJ...896...25F} from adiabatic to radiative regime including the self-absorption phase and the dimensionless factor $\xi$ which provides information on the EATS \citep{1998ApJ...493L..31P, 2000ApJ...536..195C}.

\subsection{Synchrotron scenario}\label{subsec21}

Relativistic electrons are accelerated in the FS and cooled down mainly via synchrotron emission in the presence of a comoving magnetic field $B'= \sqrt{8\pi \varepsilon_Be}$, where $e$ is the energy density and $\varepsilon_B$ the fraction of magnetic energy given in the FS. Hereafter, we use the prime and unprimed quantities to refer them in the comoving and observer frames, respectively. The acceleration process leads to that electrons with Lorentz factors ($\gamma_e$) come by a distribution of the form $N(\gamma_e)\,d\gamma_e \propto \gamma_e^{-p}\,d\gamma_e$ with $p$ the electron power index.  We consider a radially off-axis jet with an equivalent kinetic energy given by:
\bary\label{eq:ek}
E= \tilde{E}\,\Gamma^{-\alpha_s} \frac{1}{(1+ \Delta \theta^2\Gamma^2)^3}\,,
\eary
where $\tilde{E}$ is the characteristic energy, $\Delta \theta=\theta_{\rm obs} - \theta_{\rm j}$ corresponds to the viewing angle ($\theta_{\rm obs}$) and the half-opening angle of the jet ($\theta_{\rm j}$) and $\Gamma$ is the bulk Lorentz factor. We consider that the circumburst medium can be constant ($n$) or stratified (with a profile given by the stellar-wind $\propto A_{\rm W}r^{-2}$ with $A_{\rm W}$ the density parameter).


\subsubsection{Constant-density medium}

We assume an evolution of the FS with an isotropic equivalent-kinetic energy $E=\frac{4\pi}{3} m_pc^2 n r^{3}\Gamma_0^{\epsilon} \Gamma^{2-\epsilon}$ \citep[Blandford-McKee solution;][]{1976PhFl...19.1130B}, where $\epsilon=0$ corresponds to the adiabatic regime and $\epsilon=1$ to the fully radiative one, and a radial distance $r=c \xi \Gamma^2 t/(1+z)$ with $c$ the speed of light, $m_p$ is the proton mass and $z$ the redshift. Therefore, the evolution of the bulk Lorentz factor is given by: 

\bary\label{Gamma}
\Gamma= 9.8\,\left(\frac{1+z}{1.022}\right)^{\frac{3}{\delta+8-\epsilon}}\,\xi^{-\frac{6}{\delta+8-\epsilon}}\, n^{-\frac{1}{\delta+8-\epsilon}}_{-4} \, \Delta \theta^{-\frac{6}{\delta+8-\epsilon}}_{15^\circ} \,\Gamma_0^{-\frac{\epsilon}{\delta+8-\epsilon}}\, \tilde{E}^{\frac{1}{\delta+8-\epsilon}}_{52} \,t^{-\frac{3}{\delta+8-\epsilon}}_{5}\,,
\eary

with $\delta=\alpha_s+6$. Using the bulk Lorentz factor (eq.~\ref{Gamma}) and the synchrotron afterglow theory introduced in \cite{1998ApJ...497L..17S} for the fully adiabatic regime, we derive, in this formalism, the relevant quantities of synchrotron emission originated from the FS. The minimum and cooling electron Lorentz factors can be written as:

{\small
\bary\label{eLor_syn_ism}
\gamma_m&=& 32.6\,\left(\frac{1+z}{1.022}\right)^{\frac{3}{\delta+8-\epsilon}}\, \xi^{-\frac{6}{\delta+8-\epsilon}}\,g(p)\, \varepsilon_{e,-2} \, n_{-4}^{-\frac{1}{\delta+8-\epsilon}}\,\Delta\theta^{-\frac{6}{\delta+8-\epsilon}}_{15^\circ}\,\Gamma_0^{-\frac{\epsilon}{\delta+8-\epsilon}}\, \tilde{E}^{\frac{1}{\delta+8-\epsilon}}_{52}\,t^{-\frac{3}{\delta+8-\epsilon}}_{5},\cr
\gamma_c&=& 4.0\times 10^8 \left(\frac{1+z}{1.022}\right)^{\frac{\delta-1-\epsilon}{\delta+8-\epsilon}}\xi^{\frac{2(1-\delta+\epsilon)}{\delta+8-\epsilon}} (1+Y)^{-1}\, \varepsilon_{B,-4}^{-1}\,n_{-4}^{-\frac{\delta+5-\epsilon}{\delta+8-\epsilon}}\,\Delta\theta^{\frac{18}{\delta+8-\epsilon}}_{15^\circ} \,\Gamma_0^{\frac{3\epsilon}{\delta+8-\epsilon}}\,\tilde{E}^{-\frac{3}{\delta+8-\epsilon}}_{52}\,t^{\frac{1-\delta+\epsilon}{\delta+8-\epsilon}}_{5}\,,
\eary
}
respectively. Here, $Y$ is the Compton parameter, $g(p)=(p-2)/(p-1)$ whereas $\epsilon_e$ is the fraction of energy given to accelerate the electron population. Using the electron Lorentz factors (eq. \ref{eLor_syn_ism}), the characteristic and cooling spectral breaks for synchrotron radiation are
{\small
\bary\label{En_br_syn_ism}
\nu_{\rm m}&\simeq& 2.0\times 10^{-3}\,{\rm GHz}\,\, \left(\frac{1+z}{1.022}\right)^{\frac{4-\delta+\epsilon}{\delta+8-\epsilon}}\,\xi^{-\frac{24}{\delta+8-\epsilon}}\, \varepsilon^2_{e,-2}\,\varepsilon_{B,-4}^{\frac12}\,n_{-4}^{\frac{\delta-\epsilon}{2(\delta+8-\epsilon)}}\Delta\theta_{15^\circ}^{-\frac{24}{\delta+8-\epsilon}}\,\Gamma_0^{-\frac{4\epsilon}{\delta+8-\epsilon}}\,E_{52}^{\frac{4}{\delta+8-\epsilon}}\, t_{5}^{-\frac{12}{\delta+8-\epsilon}},\cr
\nu_{\rm c}&\simeq& 7.6\times 10^{4}\,{\rm keV}\,\, \left(\frac{1+z}{1.022}\right)^{\frac{\delta-4-\epsilon}{\delta+8-\epsilon}}\,\xi^{-\frac{4(2+\delta-\epsilon)}{\delta+8-\epsilon}}\, (1+Y)^{-2} \,\varepsilon_{B,-4}^{-\frac32}\,n_{-4}^{-\frac{16+3\delta-3\epsilon}{2(\delta+8-\epsilon)}}\, \Delta\theta_{15^\circ}^{\frac{24}{\delta+8-\epsilon}}\,\Gamma_0^{\frac{4\epsilon}{\delta+8-\epsilon}}\, E_{52}^{-\frac{4}{\delta+8-\epsilon}}\, t_{5}^{-\frac{2(2+\delta-\epsilon)}{\delta+8-\epsilon}}\,,
\eary
}
respectively. Considering the maximum emissivity, the total number of radiating electrons and the luminosity distance $D_{\rm z}$, the maximum flux emitted by synchrotron radiation is given by 
{\small
\bary\label{flux_syn}
F_{\rm max} &\simeq& 0.2\,{\rm mJy}\,\, \left(\frac{1+z}{1.022}\right)^{\frac{16-\delta+\epsilon}{\delta+8-\epsilon}}\,\xi^{\frac{6(\delta-\epsilon)}{\delta+8-\epsilon}}\,\varepsilon_{B,-4}^{\frac12}\,n_{-4}^{\frac{8+3\delta-3\epsilon}{2(\delta+8-\epsilon)}}\,\Delta\theta_{15^\circ}^{-\frac{48}{\delta+8-\epsilon}}\, D^{-2}_{\rm z, 26.3}\,\Gamma_0^{-\frac{8\epsilon}{\delta+8-\epsilon}}\,E_{52}^{\frac{8}{\delta+8-\epsilon}}\, t_{5}^{\frac{3(\delta-\epsilon)}{\delta+8-\epsilon}}\,.
\eary
}

The synchrotron spectral breaks in the self-absorption regime are derived from $\nu_{\rm a,1}=\nu_{\rm c}\tau^{\frac35}_{m}$, $\nu_{\rm a,2}=\nu_{\rm m}\tau^{\frac{2}{p+4}}_{m}$ and $\nu_{\rm a,3}=\nu_{\rm m}\tau^{\frac35}_{c}$ with the optical depths given by $\tau_{m}\simeq\frac{5}{3}\frac{q_e n r}{B'\gamma^5_{\rm m}}$ and $\tau_{c}\simeq\frac{5}{3}\frac{q_e n r}{B'\gamma^5_{\rm c}}$. 

The light curves in the fast cooling regime are:

{\small
\begin{eqnarray}
\label{k0_fast}
F^{\rm syn}_{\nu}\propto \begin{cases}
 t\,\nu^{\frac13},\hspace{1.5cm}
\hspace{1.2cm} {\rm for} \hspace{0.2cm} \nu < \nu_{\rm a,3}, \cr
t^{\frac{4+11(\delta-\epsilon)}{3(\delta+8-\epsilon)}}\nu^{-\frac12},
\hspace{1.3cm} {\rm for} \hspace{0.2cm} \nu_{\rm a,3} <\nu<\nu^{\rm syn}_{\rm c} ,\hspace{.1cm} \cr
t^{\frac{2(\delta-1-\epsilon)}{\delta+8-\epsilon}}\nu^{-\frac{p-1}{2}}, \hspace{1.1cm} {\rm for} \hspace{0.2cm} \nu^{\rm syn}_{\rm c}<\nu< \nu^{\rm syn}_{\rm m},\hspace{.1cm}\cr
t^{\frac{2(2-3p+\delta-\epsilon)}{\delta+8-\epsilon}}\,\nu^{-\frac{p}{2}},\hspace{0.95cm} {\rm for} \hspace{0.2cm} \nu^{\rm syn}_{\rm m}<\nu\,, \cr
\end{cases}
\end{eqnarray}
}

and in the slow cooling regime are:

{\small
\begin{eqnarray}
\label{k0_slow1}
F^{\rm syn}_{\nu}\propto \begin{cases}
 t^{\frac{2(2+\delta-\epsilon)}{\delta+8-\epsilon}}\nu^{2},\hspace{1.5cm}
\hspace{0.22cm} {\rm for} \hspace{0.2cm} \nu < \nu_{\rm a,1}, \cr
t^{\frac{4+3(\delta-\epsilon)}{\delta+8-\epsilon}}\nu^{\frac13},
\hspace{1.65cm} {\rm for} \hspace{0.2cm} \nu_{\rm a,1} <\nu<\nu^{\rm syn}_{\rm m} ,\hspace{.1cm} \cr
t^{\frac{3(2-2p+\delta-\epsilon)}{\delta+8-\epsilon}}\nu^{-\frac{p-1}{2}}, \hspace{0.7cm} {\rm for} \hspace{0.2cm} \nu^{\rm syn}_{\rm m}<\nu< \nu^{\rm syn}_{\rm c},\hspace{.1cm}\cr
t^{\frac{2(2-3p+\delta-\epsilon)}{\delta+8-\epsilon}}\,\nu^{-\frac{p}{2}},\hspace{0.95cm} {\rm for} \hspace{0.2cm} \nu^{\rm syn}_{\rm c}<\nu\,. \cr
\end{cases}
\end{eqnarray}
}

{\small
\begin{eqnarray}
\label{k0_slow2}
F^{\rm syn}_{\nu}\propto \begin{cases}
 t^{\frac{2(2+\delta-\epsilon)}{\delta+8-\epsilon}}\nu^{2},\hspace{1.5cm}
\hspace{0.23cm} {\rm for} \hspace{0.2cm} \nu < \nu^{\rm syn}_{\rm m}, \cr
t^{\frac{2(5+\delta-\epsilon)}{\delta+8-\epsilon}}\nu^{\frac52},
\hspace{1.65cm} {\rm for} \hspace{0.2cm} \nu^{\rm syn}_{\rm m} <\nu<\nu_{\rm a,2} ,\hspace{.1cm} \cr
t^{\frac{3(2-2p+\delta-\epsilon)}{\delta+8-\epsilon}}\nu^{-\frac{p-1}{2}}, \hspace{0.7cm} {\rm for} \hspace{0.2cm} \nu_{\rm a,2}<\nu< \nu^{\rm syn}_{\rm c},\hspace{.1cm}\cr
t^{\frac{2(2-3p+\delta-\epsilon)}{\delta+8-\epsilon}}\,\nu^{-\frac{p}{2}},\hspace{0.95cm} {\rm for} \hspace{0.2cm} \nu^{\rm syn}_{\rm c}<\nu\,. \cr
\end{cases}
\end{eqnarray}
}

\subsubsection{Stellar-wind medium}

In the case of a stratified stellar-wind like medium, the number density is given by $n(r)=\frac{\rho(r)}{m_p}=\frac{A}{m_p}\,r^{-2}$ where $A=\frac{\dot{M}}{4\pi\, v}=\,5\times 10^{11}\,A_{\rm W} \,{\rm g\,cm^{-1}}$, with $\dot{M}$ the mass-loss rate and $v$ the velocity of the outflow \citep[e.g., see][]{2016ApJ...818..190F}. Taking into account the Blandford-McKee solution for a stratified stellar-wind like medium, the bulk Lorentz factor derived through the adiabatic evolution \citep{1976PhFl...19.1130B, 1997ApJ...489L..37S} is given by
{\small
\bary\label{Gamma_wind}
\Gamma= 16.2 \left(\frac{1+z}{1.022}\right)^{\frac{1}{\delta+4-\epsilon}}\, \xi^{-\frac{2}{\delta+4-\epsilon}}\, A_{\rm W,-1}^{-\frac{1}{\delta+4-\epsilon}}\, \Delta \theta^{-\frac{6}{\delta+4-\epsilon}}_{15^\circ} \,\Gamma_0^{-\frac{\epsilon}{\delta+4-\epsilon}}\, \tilde{E}^{\frac{1}{\delta+4-\epsilon}}_{52} \,t^{-\frac{1}{\delta+4-\epsilon}}_{5}\,,
\eary
}
with the characteristic energy given by {\small $\tilde{E}= \frac{16\pi}{3}\, (1+z)^{-1}\,\xi^{2}\, A_{\rm W} \, \Delta \theta^6\,\Gamma_{0}^{\epsilon}\, \Gamma^{\delta+4-\epsilon}\,t\,$}.
Using the bulk Lorentz factor (eq.~\ref{Gamma_wind}) and the synchrotron afterglow theory for a wind-like medium \citep{2000ApJ...536..195C, 2000ApJ...543...66P}, we derive the relevant quantities of synchrotron emission for our model in the fully adiabatic regime. The minimum and cooling electron Lorentz factors are given by:
{\small
\bary\label{eLor_syn_wind}
\gamma_m&=& 41.5\left(\frac{1+z}{1.022}\right)^{\frac{1}{\delta+4-\epsilon}}\, \xi^{-\frac{2}{\delta+4-\epsilon}}\,g(p)\, \varepsilon_{\rm e,-2}\,\Delta\theta^{\frac{-6}{\delta+4-\epsilon}}_{15^\circ} \, A_{\rm W,-1}^{-\frac{1}{\delta+4-\epsilon}}\,\Gamma_0^{-\frac{\epsilon}{\delta+4-\epsilon}}\, \tilde{E}^{\frac{1}{\delta+4-\epsilon}}_{52}\,t^{-\frac{1}{\delta+4-\epsilon}}_{5}, \cr
\gamma_c&=& 52.1 \left(\frac{1+z}{1.022}\right)^{-\frac{\delta+3-\epsilon}{\delta+4-\epsilon}} (1+Y)^{-1}\,\xi^{\frac{2(\delta+3-\epsilon)}{\delta+4-\epsilon}}\, \varepsilon_{B,-4}^{-1}\,A_{\rm W,-1}^{-\frac{\delta+5-\epsilon}{\delta+4-\epsilon}}\,\Delta\theta^{-\frac{6}{\delta+4-\epsilon}}_{15^\circ} \,\Gamma_0^{-\frac{\epsilon}{\delta+4-\epsilon}}\,\tilde{E}^{\frac{1}{\delta+4-\epsilon}}_{52}\,t^{\frac{\delta+3-\epsilon}{\delta+4-\epsilon}}_{5}\,. 
\eary
}
The characteristic and cooling spectral breaks for synchrotron emission are:
{\small
\bary\label{En_br_syn_wind}
\nu_{\rm m}&\simeq& 1.0\times 10^{14}{\rm Hz}\, \left(\frac{1+z}{1.022}\right)^{\frac{2}{\delta+4-\epsilon}}\, \xi^{-\frac{2(\delta+6-\epsilon)}{\delta+4-\epsilon}} \,\varepsilon^2_{\rm e,-2}\,\varepsilon_{B,-4}^{\frac12}\,A_{W,-1}^{\frac{\delta-\epsilon}{2(\delta+4-\epsilon)}}\, \,\Delta\theta_{15^\circ}^{-\frac{12}{\delta+4-\epsilon}}\,\Gamma_0^{-\frac{2\epsilon}{\delta+4-\epsilon}}\, E_{52}^{\frac{2}{\delta+4-\epsilon}}\, t_{5}^{-\frac{\delta+6-\epsilon}{\delta+4-\epsilon}},\cr
\nu_{\rm c}&\simeq& 1.1\times 10^{14}{\rm Hz}\,\, \left(\frac{1+z}{1.022}\right)^{-\frac{2(\delta +3-\epsilon)}{\delta+4-\epsilon}}\,\xi^{\frac{2(\delta+2-\epsilon)}{\delta+4-\epsilon}}\,(1+Y)^{-2}\, \,\varepsilon_{B,-4}^{-\frac32} A_{\rm W,-1}^{-\frac{3\delta+16-3\epsilon}{2(\delta+4-\epsilon)}}\,\Delta\theta_{15^\circ}^{-\frac{12}{\delta+4-\epsilon}}\,\Gamma_0^{-\frac{2\epsilon}{\delta+4-\epsilon}}\,E_{52}^{\frac{2}{\delta+4-\epsilon}}\, t_{5}^{\frac{\delta+2-\epsilon}{\delta+4-\epsilon}}\,,
\eary
}
respectively. Given the maximum emissivity in a stratified stellar-wind like medium, the maximum flux radiated by synchrotron emission is given by:
{\small
\bary\label{Flux_syn_wind}
F_{\rm max} &\simeq& 1.9\times 10^{3}\,{\rm mJy}\,\, \left(\frac{1+z}{1.022}\right)^{\frac{2(\delta+5-\epsilon)}{\delta+4-\epsilon}}\, \xi^{-\frac{4}{\delta+4-\epsilon}} \,\varepsilon_{B,-4}^{\frac12}\,A_{\rm W,-1}^{\frac{3\delta+8-3\epsilon}{2(\delta+4-\epsilon)}}\, D^2_{\rm z, 26.3}\,\Delta\theta_{15^\circ}^{-\frac{12}{\delta+4-\epsilon}}\,\Gamma_0^{-\frac{2\epsilon}{\delta+4-\epsilon}}\,E_{52}^{\frac{2}{\delta+4-\epsilon}}\, t_{5}^{-\frac{2}{\delta+4-\epsilon}}\,.
\eary
}

The synchrotron spectral breaks in the self-absorption regime are derived from $\nu_{\rm a,1}=\nu_{\rm c}\tau^{\frac35}_{m}$, $\nu_{\rm a,2}=\nu_{\rm m}\tau^{\frac{2}{p+4}}_{m}$ and $\nu_{\rm a,3}=\nu_{\rm m}\tau^{\frac35}_{c}$ with the optical depths given by $\tau_{m}\propto \frac{q_e A_{\rm W} r^{-1}}{B'\gamma^5_{\rm m}}$ and $\tau_{c} \propto \frac{q_e A_{\rm W} r^{-1}}{B'\gamma^5_{\rm c}}$. 

The light curves in the fast cooling regime are:

{\small
\begin{eqnarray}
\label{k2_fast}
F^{\rm syn}_{\nu}\propto \begin{cases}
 t^{\frac{8+3(\delta-\epsilon)}{\delta+4-\epsilon}}\nu^{\frac13},\hspace{1.5cm}
\hspace{0.35cm} {\rm for} \hspace{0.2cm} \nu < \nu_{\rm a,3}, \cr
t^{\frac{\epsilon-\delta-8}{3(\delta+4-\epsilon)}}\nu^{-\frac12},
\hspace{1.64cm} {\rm for} \hspace{0.2cm} \nu_{\rm a,3} <\nu<\nu^{\rm syn}_{\rm c} ,\hspace{.1cm} \cr
t^{\frac{\delta-2-\epsilon}{2(\delta+4-\epsilon)}}\nu^{-\frac{p-1}{2}}, \hspace{1.3cm} {\rm for} \hspace{0.2cm} \nu^{\rm syn}_{\rm c}<\nu< \nu^{\rm syn}_{\rm m},\hspace{.1cm}\cr
t^{\frac{2(2-3p)+(\epsilon-\delta)(p-2)}{2(\delta+4-\epsilon)}}\,\nu^{-\frac{p}{2}},\hspace{0.3cm} {\rm for} \hspace{0.2cm} \nu^{\rm syn}_{\rm m}<\nu\,, \cr
\end{cases}
\end{eqnarray}
}

whereas in the slow cooling regime are:

{\small
\begin{eqnarray}
\label{k2_slow1}
F^{\rm syn}_{\nu}\propto \begin{cases}
 t^{\frac{2(2+\delta-\epsilon)}{\delta+4-\epsilon}}\nu^{2},\hspace{1.5cm}
\hspace{0.72cm} {\rm for} \hspace{0.2cm} \nu < \nu_{\rm a,1}, \cr
t^{\frac{\delta-\epsilon}{3(\delta+4-\epsilon)}}\nu^{\frac13},
\hspace{2.15cm} {\rm for} \hspace{0.2cm} \nu_{\rm a,1} <\nu<\nu^{\rm syn}_{\rm m} ,\hspace{.1cm} \cr
t^{\frac{2+\delta-\epsilon-p(6+\delta-\epsilon)}{2(\delta+4-\epsilon)}}\nu^{-\frac{p-1}{2}}, \hspace{0.7cm} {\rm for} \hspace{0.2cm} \nu^{\rm syn}_{\rm m}<\nu< \nu^{\rm syn}_{\rm c},\hspace{.1cm}\cr
t^{\frac{2(2-3p)+(\epsilon-\delta)(p-2)}{2(\delta+4-\epsilon)}}\,\nu^{-\frac{p}{2}},\hspace{0.63cm} {\rm for} \hspace{0.2cm} \nu^{\rm syn}_{\rm c}<\nu\,. \cr
\end{cases}
\end{eqnarray}
}

{\small
\begin{eqnarray}
\label{k2_slow2}
F^{\rm syn}_{\nu}\propto \begin{cases}
 t^{\frac{2(2+\delta-\epsilon)}{\delta+4-\epsilon}}\nu^{2},\hspace{1.5cm}
\hspace{0.43cm} {\rm for} \hspace{0.2cm} \nu < \nu^{\rm syn}_{\rm m}, \cr
t^{\frac{14+5(\delta-\epsilon)}{2(\delta+4-\epsilon)}}\nu^{\frac52},
\hspace{1.75cm} {\rm for} \hspace{0.2cm} \nu^{\rm syn}_{\rm m} <\nu<\nu_{\rm a,2} ,\hspace{.1cm} \cr
t^{\frac{2+\delta-\epsilon-p(6+\delta-\epsilon)}{2(\delta+4-\epsilon)}}\nu^{-\frac{p-1}{2}}, \hspace{0.43cm} {\rm for} \hspace{0.2cm} \nu_{\rm a,2}<\nu< \nu^{\rm syn}_{\rm c},\hspace{.1cm}\cr
t^{\frac{2(2-3p)+(\epsilon-\delta)(p-2)}{2(\delta+4-\epsilon)}}\,\nu^{-\frac{p}{2}},\hspace{0.35cm} {\rm for} \hspace{0.2cm} \nu^{\rm syn}_{\rm c}<\nu\,. \cr
\end{cases}
\end{eqnarray}
}

\section{Polarization model}\label{sec3}

Since 1999, the phenomena of polarization, the confinement of wave vibrations to a certain geometrical direction, has been detected in GRBs \citep{1999A&A...348L...1C}. Further studies indicate that the polarization degree ($\Pi$) can have high variability, but the polarization angle (P.A.; $\theta_p$) remains roughly the same, for an observer outside the jet \citep{2021MNRAS.507.5340T}.
Polarization is commonly attributed to synchrotron radiation behind shock waves. This makes it dependent on the magnetic field configuration and the geometry of the shock, as they will define the the P.D. on each point and its integration over the whole image \citep{Gill-1}. 
The Stokes parameters (I, Q, U, and V) control the approach to polarization calculation, and normally only linear polarization is considered.
From this point on, we refer to the observer and comoving frames as unprimed and primed, respectively.
The stokes parameters are expressed as

\begin{eqnarray}
    &V = 0, \hspace{2cm} &\theta_p = \frac{1}{2}\arctan{\frac{U}{Q}},\,\cr
    &\frac{U}{I} = \Pi'\sin{2\theta_p}, \hspace{1cm}
    &\frac{Q}{I} = \Pi'\cos{2\theta_p}.
\end{eqnarray}

And the measured stokes parameters are the sum over the flux \citep{Granot-P2}, so
 
\begin{eqnarray}
       &\frac{U}{I} = \frac{\int \mathrm{d}F_\nu\Pi'\sin{2\theta_p}}{\int \mathrm{d}F_\nu},\hspace{1cm}
    \frac{Q}{I} =  \frac{\int \mathrm{d}F_\nu\Pi'\cos{2\theta_p}}{\int \mathrm{d}F_\nu},  \\
    &\Pi = \frac{\sqrt{Q^2+U^2}}{I}.
\end{eqnarray}

The relationship $\mathrm{d}F_\nu \propto   \delta_D^3L'_{\nu'}\mathrm{d}\Omega$ -- where $L'_{\nu'}$ is the spectral luminosity and $\mathrm{d}\Omega$ is the element of solid angle of the fluid element in relation to the source -- allows the introduction of the factors regarding the geometry of the magnetic field and outflow by using \citep{ribicky}
\begin{eqnarray}
    L'_{\nu'} \propto (\nu')^{-\alpha} (\sin{\chi'})^\epsilon r^m \propto (\nu')^{-\alpha} (1-\hat{n}' \cdot \hat{B}')^{\epsilon/2} r^m.
\end{eqnarray}

The parameter $\chi$ is the angle between the local magnetic field and the particle's direction of motion, and due to the highly beamed nature of synchrotron emission, this angle is also the pitch angle. The geometrical considerations of polarization can then be taken by averaging this factor over the local probability distribution of the magnetic field \citep[see Eq. 15 of][]{Gill-1},

\begin{eqnarray}
    \Lambda = \expval{(1-\hat{n}' \cdot \hat{B}')^{\epsilon/2}}.
\end{eqnarray}

It is possible to do a Lorentz transformation on the unit vectors, like $\hat{n}$, or a certain configuration of $\hat{B}$ to express $\Lambda$ in terms of different magnetic field configurations \citep{Gill-1, Lyutikov, Granot-P2}:

\begin{eqnarray}
    \Lambda_{\rm ord} &\approx& \left[\left(\frac{1-\Tilde{\xi}}{1+\Tilde{\xi}}\right)\cos^2{\varphi_B} + \sin^2{\varphi_B}\right]^{\epsilon/2}\label{eq:pd_ord},\\
    \Lambda_{\perp} &\approx& \expval{\Lambda_{ord}(\Tilde{\xi}, \varphi_B)}_{\varphi_B}\label{eq:pd_perp},\\
    \Lambda_{\parallel} &\approx& \left[\frac{\sqrt{4\Tilde{\xi}}}{1+\Tilde{\xi}}\right]^\epsilon \label{eq:pd_par},
\end{eqnarray}

where $\varphi_B$ as the azimuthal angle of the magnetic field measured from a reference point. $\tilde{\xi} \equiv (\Gamma\tilde{\theta})^2$, taking in consideration the approximations of $\Tilde{\mu} = \cos{\tilde{\theta}} \approx 1 - \tilde{\theta}^2/2$ and $\beta \approx 1 - 1/2\Gamma^2 $, which leads to $ \delta_D \approx \frac{2\Gamma}{1+\Tilde{\xi}}$ where $\tilde{\theta}$ the polar angle measured from the Line of Sight (LOS).

One of the still-unsolved mysteries of GRBs is the configuration of the magnetic field present at different regions of emission. As such, various possible configurations must be explored in a topic where magnetic field geometry is of paramount relevance, like polarization.
The considerations regarding the magnetic field geometry are varied based on the GRB epoch of relevance. For a scenario where the afterglow is described by a FS, two of the most suitable configurations are: a random perpendicular configuration -- where the anisotropy factor $b \equiv \frac{2\expval{B_\parallel^2}}{\expval{B_\perp^2}}= 0$ -- confined to the shock plane; and an ordered configuration parallel to the velocity vector, where $b\rightarrow\infty$. More complex configurations with multi-component, where the anisotropy is $b>0$, magnetic fields have been done \citep{2020MNRAS.491.5815G, 2021MNRAS.507.5340T, 2020ApJ...892..131S, 2018ApJ...861L..10C}, as it is warranted and needed, however, for the purposes of this paper we limit ourselves to the two following cases.

\paragraph*{Random magnetic field ($B_\perp,\,b=0$)}
In this scenario, the symmetry of the random magnetic field configuration, perpendicular to the shock plane, causes the polarization over the image to disappear when if the beaming cone is wholly contained within the jet aperture or if it is seen along the axis ($\theta_{\rm obs} = 0$). To break the symmetry, the jet must be viewed close to its edge ($q\equiv \frac{\theta_{\rm obs}}{\theta_j} \gtrsim 1+\xi_j^{-1/2}$), where missing emission (from $\theta > \theta_j$) results only in partial cancellation \citep{Waxman}.
The equation necessary to calculate this polarization is explicitly laid out as Eq. 5 in \citep{Granot-P2}.


\paragraph*{Ordered magnetic field ($B_\parallel, \,b\rightarrow\infty$)}
For the ordered magnetic field, a configuration parallel to the velocity vector, the same symmetry observations hold true and the calculation follows \citep{Granot-P2, Gill-1}, with $\Lambda(\tilxi) = \Lambda_\parallel$ from \Cref{eq:pd_par}. 

By substituting the following integration limits

\begin{eqnarray}\label{conections}
    \cos{\psi(\tilxi)} = \frac{(1-q)^2 \xi_j - \tilxi}{2q\sqrt{\xi_j\tilxi }}, \qquad \xi_j = (\Gamma\theta_j)^2, \qquad \xi_\pm = (1\pm q)^2\xi_j,
\end{eqnarray}

with an appropriate prescription of the bulk Lorentz factor $\Gamma(t)$, the evolution of the opening angle of the jet $\theta_j(t)$, and the parameters required to describe these expressions as described in Section \ref{sec2} and \cite{2020ApJ...896...25F}, we can obtain the temporal evolution of polarization.

\subsection{Polarization evolution for a Forward-Shock } 
\begin{figure}
{\includegraphics[width=\textwidth]{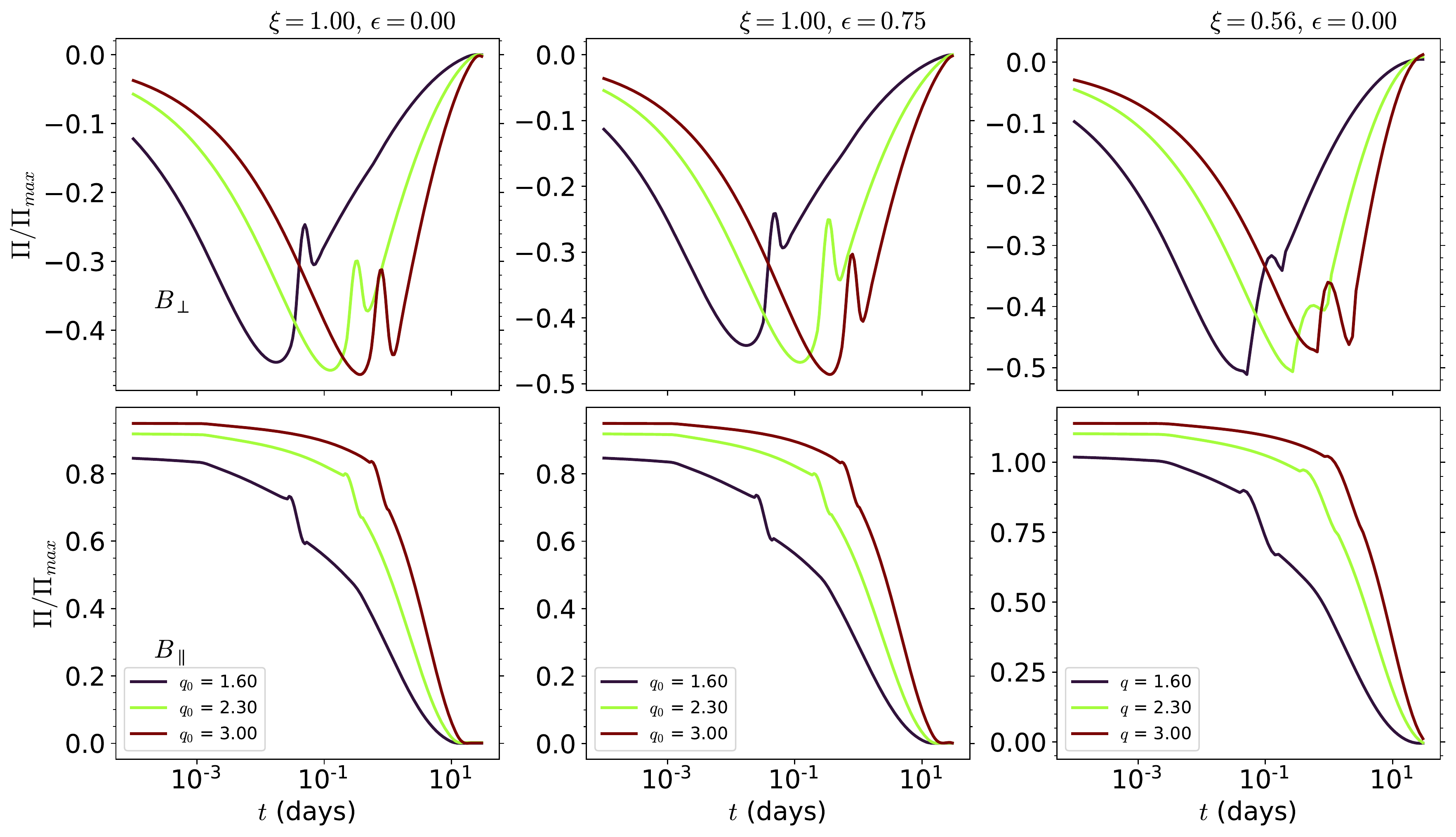}}
\caption{Polarization curves for our fiducial model, considering a constant-medium. The top row shows the perpendicular magnetic field configuration, the bottom row shows the parallel one. Each column represent a different pairing $\xi$ and $\epsilon$.}
\label{fig:general_k0}
\end{figure}

\begin{figure}
{\includegraphics[width=\textwidth]{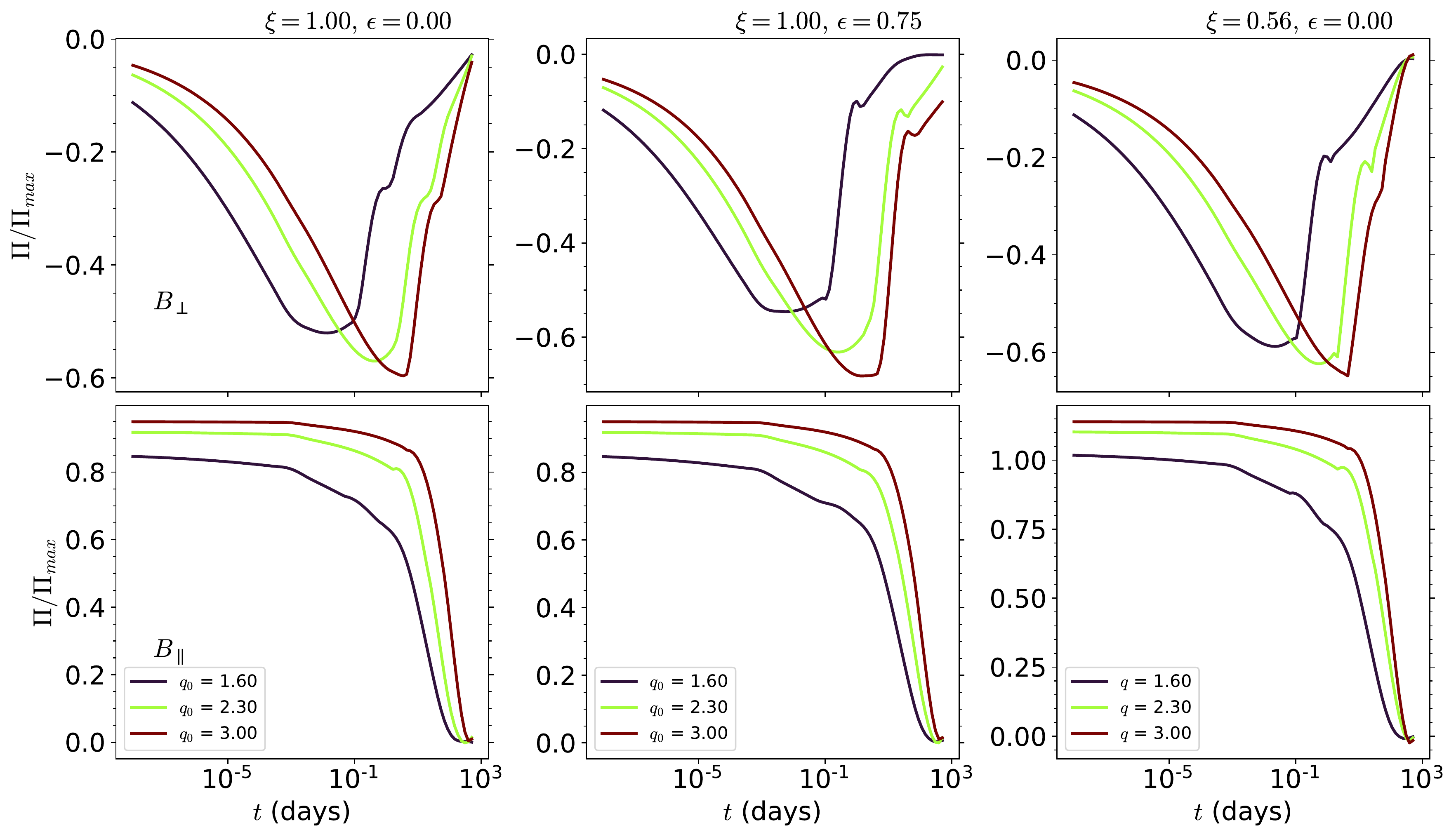}}
\caption{Same as \Cref{fig:general_k0}, but for a wind-like medium.} 
\label{fig:general_k2}
\end{figure}

\Cref{fig:general_k0,fig:general_k2} show the temporal evolution of polarization degree for our chosen magnetic field configurations in two distinct scenarios regarding the density of the circumburst medium -- here considered a constant density and a wind-like medium. Each column of these figures represents a chosen combination of the $\epsilon$ and $\xi$ parameters. Table \ref{tab:table_general} shows the values required to generate \Cref{fig:general_k0,fig:general_k2}. We highlight that the generic values were determined based on the typical range reported for each parameter in the GRB synchrotron literature \citep[for reviews, see][]{Kumar, 2014ARA&A..52...43B}. The values of observation angle are varied over a range between $8$ and $15\deg$.\footnote{Over the course of this manuscript we will be using $\deg$ as the abbreviation of degree.} This range of values is shown in these figures with different colored lines, each one standing for a value of $q_0=\frac{\theta_{\rm obs}}{\theta_{\rm j,0}}$, the ratio between the observation angle and initial opening angle of the jet.

The synchrotron model chosen is a homogeneous off-axis jet, in which the equivalent kinetic energy is parameterized with a power-law velocity distribution (see \Cref{eq:ek}), that suffers sideways expansion (SE) with the comoving speed of sound given as Eq. 10 in \cite{2000ApJ...543...90H}. The homogeneous jet case has been studied by a few works now, such as \cite{2002ApJ...570L..61G, 2004MNRAS.354...86R}, however, these works only have explored random magnetic field on a fully adiabatic regime in a constant medium. Nonetheless, comparing results with the leftmost column of \Cref{fig:general_k0}, we see the typical double peak behavior for a homogeneous SE jet, reported by \cite{2004MNRAS.354...86R} for $q_0<5$, is presented for us as well. Some discrepancies are shown, with our polarization being initially higher at early times (increasingly so as $q_0 \rightarrow 1$) and overall in the magnitude of the peaks. The highest likelihood culprit for these differences is the choice of synchrotron model and parameter values. The center column of \Cref{fig:general_k0} presents the case for a partially radiative scenario, and it behaves quite similarly to the adiabatic case, with only a change in magnitude of the peaks being observable. The deceleration of the relativistic outflow by the circumburst medium is faster when it lies in the radiative regime rather than adiabatic one, and the temporal evolution of polarization is modified \citep{2000ApJ...532..281B, 2005ApJ...619..968W}. For our model, this has resulted in an enhancement of the increase in polarization as $q_0$ grows, but smaller second peaks. The rightmost side of \Cref{fig:general_k0} displays the case for an adiabatic regime with $\xi = 0.56$ \citep{2000ApJ...536..195C}. The variation on $\xi$ causes the emission to arrive earlier or later, and this produces a difference in the magnitudes of the peaks, as observed in \Cref{fig:general_k0} \citep{1997ApJ...491L..19W, 2000ApJ...536..195C, 1998ApJ...493L..31P}. The polarization behaviour flips in comparison with $\xi=1$, with the peak increasing as $q_0 \rightarrow 1$; comparatively, the second peak remains mostly the same. The parallel case presents similar behavior for all three considered cases. A small change is observed at the sharpness of decline of polarization at jet-break (where the synchrotron model bulk Lorentz factor changes regime to follow the on-axis calculations presented by \cite{2020ApJ...896...25F}) and post-break, with a stronger discontinuity happening with a decreasing value of $\xi$.

\Cref{fig:general_k2} shows the polarization evolution for the wind-like medium. \cite{2004A&A...422..121L} expected that the polarization evolved slower for a wind-like medium, as the relationship between afterglow timescale and density was $t\propto (E/n)^{\frac{1}{(3-k)}}$ \citep[][and with $k=2$ for a wind-like medium]{Kumar, 2022arXiv220502459F}, and this is observed here too. For a convenience of observation, the limits of the timescale have been expanded. Other significant differences between the constant-density medium and this scenario are the higher initial polarization peak and lower magnitude of the second peak, in all likelihood due to the lower value of bulk Lorentz factor at later times. Between the chosen values of $\xi$ and $\epsilon$, we see that a lower value of $\xi$ increases the magnitude of the first peak while decreasing the magnitude of the second one. This is similar to the constant-density medium case, with the addendum that the second peak is reduced further when compare to $\xi=1$. For the partially radiative case, the first polarization peak is similar to the adiabatic case with higher magnitude, but the second polarization peak is further reduced.

\section{Polarization from GRB Off-axis Afterglows}\label{sec4}

In this section, we describe the polarization for a group of GRBs that show similar characteristics on their afterglow: GRB 080503, GRB 140903A, GRB 150101B, GRB 160821B, and GRB 170817A. In \cite{2020ApJ...896...25F}, the authors have explored the similarities between those bursts. We use the parameter values obtained by \cite{2020ApJ...896...25F} via Markov Chain Monte Carlo (MCMC) simulations to calculate polarization. For this section, we will adopt the notation $f(q_0=x^{\pm y}_{\pm z}) = a^{\pm b}_{\pm c}$ when the chosen values of $q_0$ result in significant differentiation on polarization or peak time.

\begin{figure}
{\includegraphics[width=\textwidth]{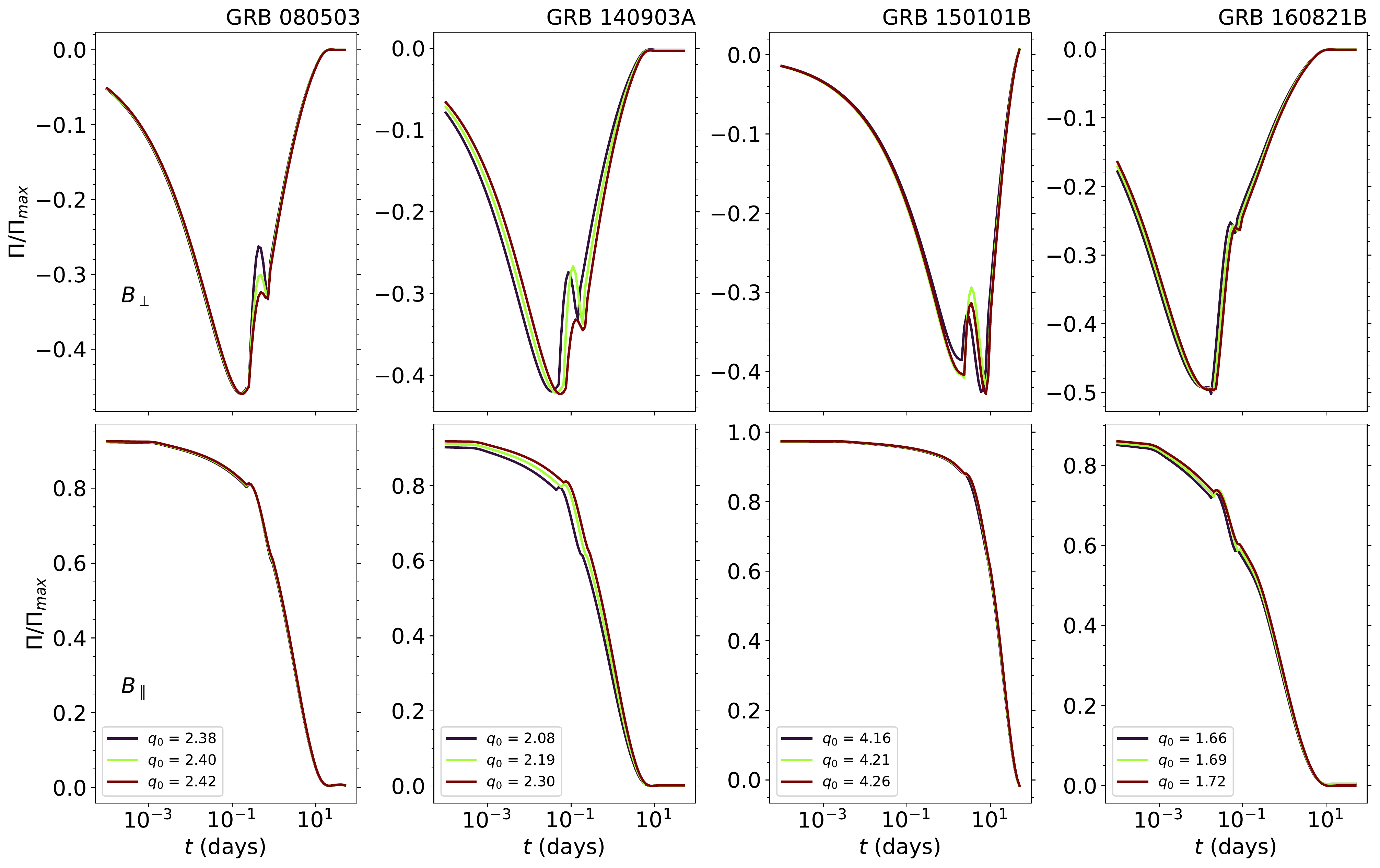}}
\caption{Expected temporal evolution of the polarization for the bursts GRB 080503, GRB 140903, GRB 150101B, GRB 160821B - respectively, from left to right -- for the parameters presented in \Cref{tab:par_mcmc}. The uncertainty of the observation angle, $\theta_{\rm obs}$, was used to return a range of values for the fraction $q$, represented by the colormap legend on the figure. All polarizations are calculated for the case $k=0$.}
\label{fig:joined_GRB}
\end{figure}

\paragraph{GRB 080503}
The first column in \Cref{fig:joined_GRB} shows the expected polarization evolution calculated for GRB 080503 for our two configurations. The parameters for calculating this polarization are presented in the first row of \Cref{tab:pol_grbs}. The granular increment of $q_0$ shows little effect in the polarization curves in either configuration, with the only major difference being the magnitude of the minimum located between peaks. The initial polarization for the time-frame we have chosen is $\abs{\Pi/\Pi_{\rm max}}\approx 5\%$ and $\Pi/\Pi_{\rm max}\approx 92\%$ for $B_\perp$ and $B_\parallel$, respectively. For $B_\perp$, the polarization evolves towards a peak of $\abs{\Pi/\Pi_{\rm max}}\approx 46\%$, with a second peak of $\abs{\Pi/\Pi_{\rm max}}\approx 33\%$ at $\sim 0.3$ and $\sim 0.8$ days, respectively. For the parallel configuration, the initial polarization decreases softly during the off-axis period by roughly $10\%$. After the jet break, the polarization drops sharply, and zero polarization is reached at $\sim 17 $ days.

\paragraph{GRB 140903A}
The second column in \Cref{fig:joined_GRB} shows the expected polarization evolution estimated for GRB 140903A. The parameters for calculating this polarization are presented in the second row of \Cref{tab:pol_grbs}. For this burst, a slightly higher degree of influence of $q_0$ is observed. The initial polarization values are $\abs{\Pi/\Pi_{\rm max}}(q_0=2.19^{+0.11}_{-0.11})\approx 7.1^{-0.6}_{+0.7}\%$ and $\Pi/\Pi_{\rm max}(q_0=2.19^{+0.11}_{-0.11})\approx 91^{+0.7}_{-1.0}\%$ for $B_\perp$ and $B_\parallel$, respectively. For the perpendicular field, the peak of $\abs{\Pi/\Pi_{\rm max}}\approx 42^{+0.4}_{-0.2}\%$ is seen at $t_{\rm peak}(q_0=2.19^{+0.11}_{-0.11}) \approx 4.4^{+0.6}_{-1.0}\times 10^{-2}$ days with a second peak of $\abs{\Pi/\Pi_{\rm max}}\approx 33^{+2}_{-1}\%$ at $t_{\rm peak} \approx 1.9^{+0.2}_{-0.5}\times 10^{-1}$ days, respectively. For the parallel configuration, the polarization at the break is $\Pi/\Pi_{\rm max}\approx 79^{+1.0}_{-0.8}\%$, and zero is achieved roughly at the same time of $\sim 6$ days.

\paragraph{GRB 150101B}

The third column in \Cref{fig:joined_GRB} shows the expected polarization calculated for GRB 150101B. The parameters for calculating this polarization are presented in the third row of \Cref{tab:pol_grbs}. The higher value of $q_0$ makes so the minute variation of the chosen values has little influence on the polarization. The initial values of polarization are $\abs{\Pi/\Pi_{\rm max}}\approx 1.5\%$ and $\Pi/\Pi_{\rm max}\approx 97\%$, for $B_\perp$ and $B_\parallel$, respectively. For $B_\perp$, the first polarization peak is $\abs{\Pi/\Pi_{\rm max}}(q_0=4.21^{+0.05}_{-0.05})\approx 40^{+0.5}_{-2.0}\%$, and the second $\abs{\Pi/\Pi_{\rm max}}\approx 42\%$ at $\sim2$ and $\sim 7$ days, respectively. For the parallel configuration, the polarization decreases by $\sim 13\%$ until the break is achieved and decreases to zero rapidly, reaching it at $\sim 40$ days.

\paragraph{GRB 160821B}
The fourth column in \Cref{fig:joined_GRB} shows the expected evolution of polarization calculated for GRB 160821B. The parameters for calculating this polarization are presented in the fourth row of \Cref{tab:pol_grbs}. The initial polarization values are $\abs{\Pi/\Pi_{\rm max}}(q_0=1.69^{+0.03}_{-0.03})\approx 17^{-0.6}_{+0.8}\%$ and $\Pi/\Pi_{\rm max}(q_0=1.69^{+0.03}_{-0.03})\approx 85.5^{+0.5}_{-0.5}\%$ for $B_\perp$ and $B_\parallel$, respectively. For the perpendicular case, the peak of $\abs{\Pi/\Pi_{\rm max}}\approx 50\%$ is seen at $t_{\rm peak}(q_0=1.69^{+0.03}_{-0.03}) \approx 2.0^{+0.2}_{-0.3}\times 10^{-2}$ days with a second peak of $\abs{\Pi/\Pi_{\rm max}}\approx 26\%$ at $t_{\rm peak} \approx 7.5^{+1.1}_{-0.9}\times 10^{-2}$ days. For the parallel configuration, the polarization at the break is $\Pi/\Pi_{\rm max}\approx 73^{+1.0}_{-0.8}\%$, and zero is achieved roughly at the same time of $\sim 6$ days.

\paragraph{GRB 170817A}

\begin{figure}
{\includegraphics[width=\textwidth]{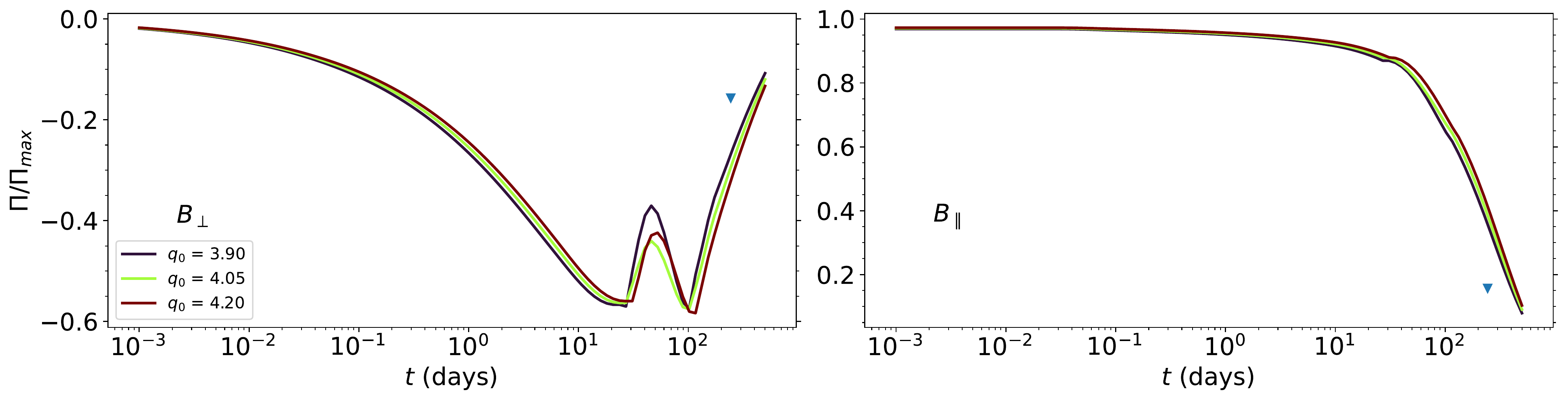}}
\caption{Expected Temporal evolution of the polarization for GRB 170817A for two configurations of magnetic fields - Perpendicular ($B_\perp$) and Parallel ($B_\parallel$). These polarization curves were calculated using the best fit values presented in Table 5 of \protect\cite{2019ApJ...871..123F}: $\tilde{E}\approx0.7 \times 10^{51}$erg, $n\approx1.0 \times 10^{-4}\, {\rm cm}^{-3}$, and $\theta_j=5.0$ deg. For $\theta_{\rm obs}$, we have used the range of $20.5\pm0.5$ deg with three values linearly spaced between the limits. The inverted triangles represent the polarization upper limits $\abs{\Pi}=12\%$ \protect\citep[derived by][]{2018ApJ...861L..10C}, re-scaled by our chosen $\Pi_{\rm max}=70\%$.}
\label{fig:joined_170817A}
\end{figure}

\Cref{fig:joined_170817A} shows the expected polarization, calculated with our model, for the different configurations of magnetic fields. GRB 170817A has been modelled by a variety of different synchrotron scenarios, while the more traditional top-hat off-axis jet has been disfavored, other models such as radially stratified ejecta \citep{2018Natur.554..207M, 2018ApJ...867...95H, 2019ApJ...871..123F}, and structured jets \citep{2017Sci...358.1559K, 2017MNRAS.472.4953L, 2018PhRvL.120x1103L} can properly describe the multiwavelength afterglow observations. One thing to note is that for the period starting two weeks after the burst, the flux can be described by a relativistic collimated jet \citep[see references above and][]{2019ApJ...884...71F}. As such, the angular structure of the jet is less relevant regarding whether (or not) a homogeneous jet can successfully describe the late afterglow.
We use the phenomenological model presented in this paper for a constant-density medium with $\xi=1$ and $\epsilon=0$ to obtain the polarization. These conditions reduce our model to the one used in \cite{2019ApJ...871..123F}, where the authors have fitted the synchrotron light curves. We have used the values reported in Table 3 in \cite{2019ApJ...871..123F} to generate the polarization curves. Based on these conditions, the polarization presents a similar behavior as the left side of \Cref{fig:general_k0}.

For the perpendicular configuration of magnetic field, we observe an initial $\sim 1.8\%$ polarization for all values of $q_0$. Then, the polarization begins its evolution towards a maximum of $\abs{\Pi/\Pi_{\rm max}}(q_0=4.05^{+0.15}_{-0.15})\approx 55^{-1}_{+1}\%$ at $t\approx 27$ days, with a second peak of $\abs{\Pi/\Pi_{\rm max}}\approx 57.5^{+0.5}_{-0.5}\%$ at $t\approx 100$ days. 
The parallel configuration has an initially high degree of polarization across the board and low influence of $q_0$, with $\Pi/\Pi_{\rm max} \approx 97\%$, and $\Pi/\Pi_{\rm max} \approx 84\%$ at the break. The blue inverted triangles \Cref{fig:joined_170817A} show the upper limits, of $\abs{\Pi} \approx 12\%$ at $t\approx 243 $ days \citep[derived by][]{2018ApJ...861L..10C}, normalized by our arbitrarily chosen value of $\Pi_{\rm max} =70\%$. Upper limits are broken by the polarization curves, with $\abs{\Pi/\Pi_{\rm max}} (B_\perp) \approx 25\%$ and $\Pi/\Pi_{\rm max} (B_\parallel) \approx 33\%$. This indicates that the chosen configurations cannot successfully describe the polarization observed for GRB 170817A.
Several attempts at constraining the magnetic field configuration of GRB 170817A have been performed \citep[e.g., see][]{2018MNRAS.478.4128G, 2020ApJ...892..131S, 2020MNRAS.491.5815G, 2021MNRAS.507.5340T}, using the available polarimetric upper limits and multiple types of outflows. These works agree that a configuration with $b=0$ ($b\rightarrow\infty$) is ruled out. An exception is the case of a wide-angled quasi-spherical outflow with energy injection, calculated by \cite{2018MNRAS.478.4128G}, which does not break the upper limits. However, this particular model is disfavoured to describe the afterglow flux of the burst. \cite{2021MNRAS.507.5340T} and \cite{2020MNRAS.491.5815G} have constrained the anisotropy of the magnetic fields to a dominant perpendicular component with a sub-dominant parallel component ($0.85\lesssim b \lesssim1.16$ and $0.66 \lesssim b \lesssim 1.49$, for each paper, respectively). 
More observations on a shorter post-burst period would be needed to constrain the magnetic field configuration further, and proper modeling of the afterglow light curve is necessary for breaking the degeneracy between models. Unfortunately, there were no polarization observations at any other frequency and time \citep{2018ApJ...861L..10C}.

\section{The closest sGRBs detected by Swift satellite}\label{sec5}

\cite{2020MNRAS.492.5011D} presented a systematic search for nearby sGRBs with similar features to GRB 170817A in the Swift database, covering 14 years of operations. A subset of four potential candidates: GRB 050906, GRB 070810B, GRB 080121, and GRB 100216A, were found between 100 and 200 ${\rm Mpc}$. These candidates were used to constrain the range of properties for X-ray counterparts of a merger of two NSs, and derived optical upper limits on the onset of a ``blue" KN, implying a low amount of lanthanide-poor ejecta (see Section 3.2 from \cite{2020MNRAS.492.5011D} and references therein).\\

\subsection{Light curves and Polarization}

\begin{figure}\label{GRBs_Swift}
{\centering
\resizebox*{\textwidth}{0.45\textheight}
{\includegraphics{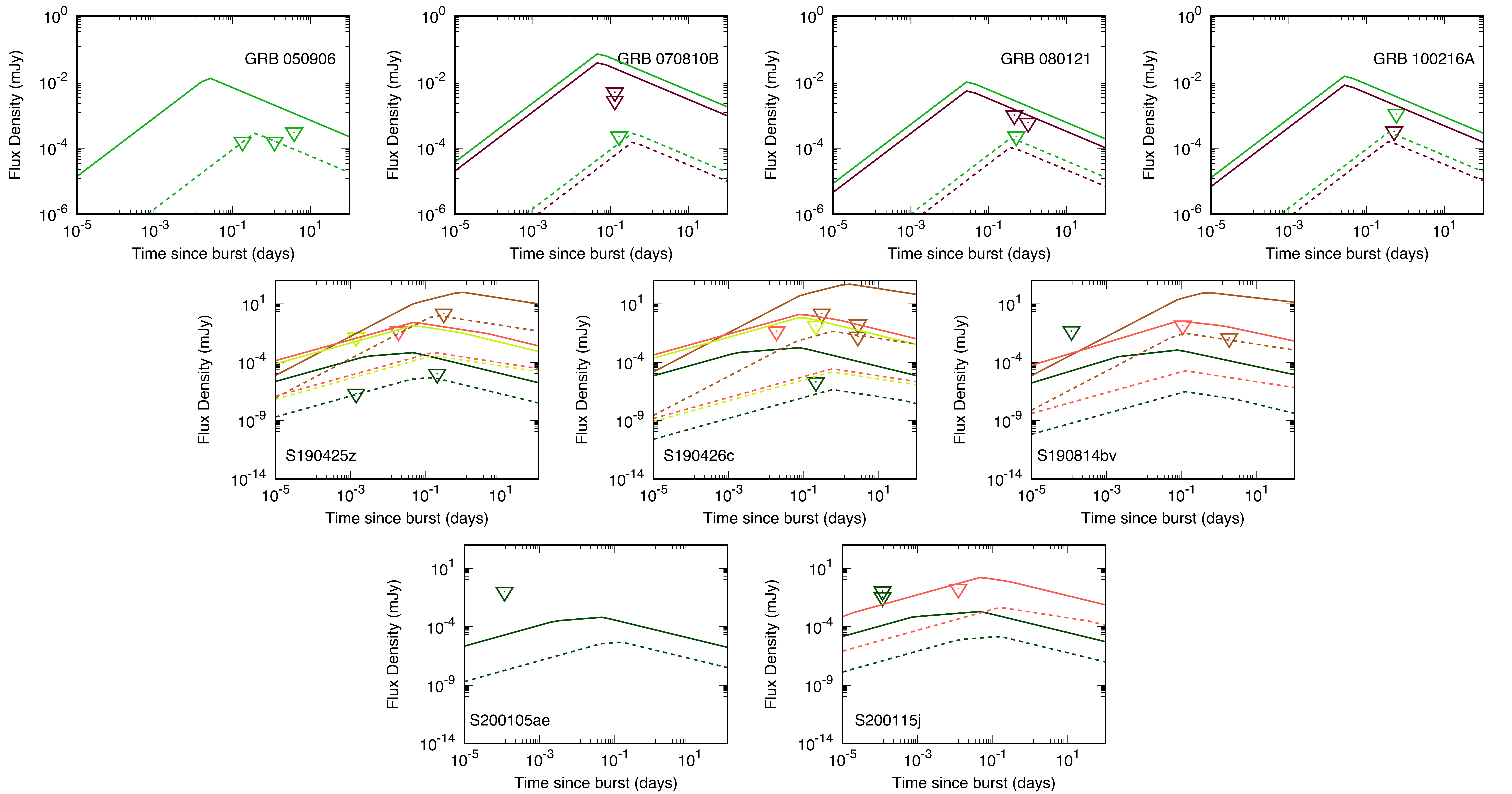}}
} 
\caption{Upper limits for GRB 050906, GRB 070810B, GRB 080121, and GRB 100216A \protect\citep[top row, taken from][]{2020MNRAS.492.5011D} and the set of GW events that could produce an electromagnetic counterpart \citep[middle and bottom rows, taken from][]{2021arXiv211103606T}. The dashed and solid lines correspond to synchrotron light curves evolving in a constant-density medium, where the solid lines represent the value of $\theta_{\rm obs}=4\,{\rm deg} (6\,{\rm deg})$ and the dashed ones use $\theta_{\rm obs}=15\,{\rm deg} (17\,{\rm deg})$. The parameter values used are presented in \Cref{tab:Best-Fit-Parameters}.}
\end{figure}

Figure \ref{GRBs_Swift} presents three rows, where the first one corresponds to a set of Swift-identified bursts between 100 and 200 Mpc. Each panel in the row shows the optical upper limits with the synchrotron light curves expected from an off-axis jet decelerating in a constant-density medium for two different viewing angles; $\theta_{\rm obs}= 4\,$ (solid lines) and $15\,{\rm deg}$ (dashed lines). The synchrotron light curves are presented at the R-band (red) and the u-band (green). The parameter values are reported in Table \ref{tab:Best-Fit-Parameters} with $\Gamma=100$, $\varepsilon_{\rm e}=0.3$, $p=2.5$, $\zeta_e=1.0$ and $\varepsilon_{\rm B}=10^{-4}$. For the chosen afterglow values, higher viewing angles (more than $15\, {\rm deg}$) are favored by our model.

\begin{figure}
{\includegraphics[width=\textwidth]{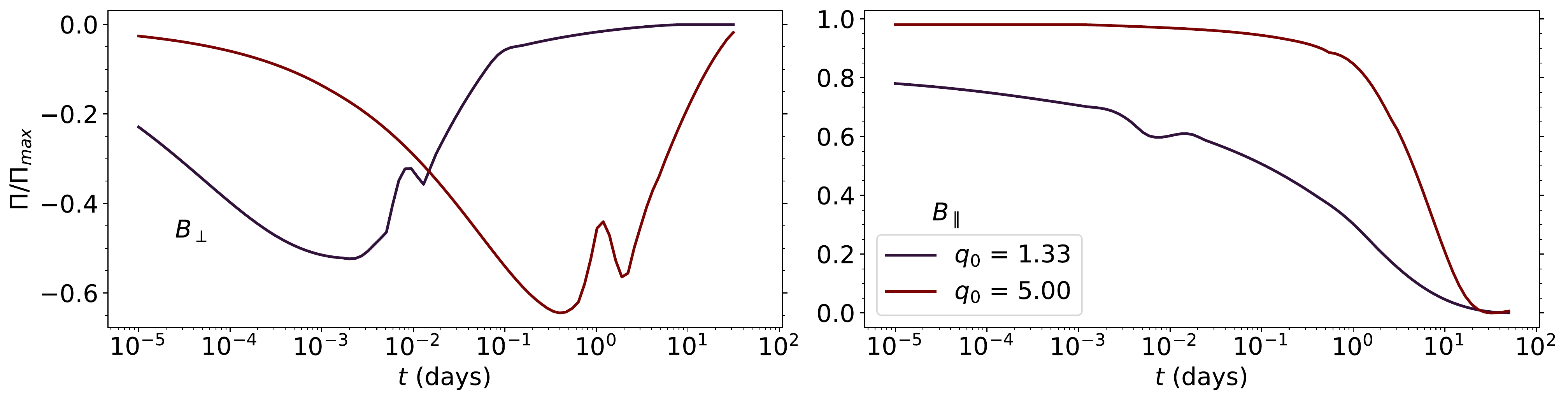}}
\caption{Expected temporal polarization evolution for the Swift-detected short GRBs located between 100 and 200 Mpc. The parameters used to calculate these polarization curves are presented in \Cref{tab:Best-Fit-Parameters}. }
\label{fig:joined_swift}
\end{figure}

\Cref{fig:joined_swift} shows the expected polarization curves that could be present on account of the parameters used to obtain Figure \ref{GRBs_Swift}. We have the perpendicular and parallel configurations presented from left to right. Two values of $q_0$ were used for these calculations, and two curves were calculated based on the different angles constrained by fitting the upper flux limits. We can notice that the set of parameters for an observation angle of $\theta_{\rm obs}=4\deg$ violates the optical upper limits. As such, we will call this set of parameters ``disallowed", and the set for which the flux is below the upper limits, with $\theta_{\rm obs}=15\deg$ as ``allowed".

Looking at the perpendicular configuration, the disallowed set presents an initially high polarization ($\abs{\Pi/\Pi_{\rm max}} =22\%$) compared to the allowed set ($\abs{\Pi/\Pi_{\rm max}} =2\%$). The peak polarization also happens earlier for the disallowed set and reaches zero earlier. Considering both evolutions, rough limits can be set for these orphan afterglows of similar characteristics. The intersection between curves would set so that the polarization must be $\abs{\Pi/\Pi_{\rm max}} < 31\%$ for $t\approx1.5\times10^{-2}$. However, since the disallowed set decreases past this point (while the allowed set increases), the requirement is that $\abs{\Pi/\Pi_{\rm max}} > 31\%$ for $t > 1.5\times10^{-2}\,{\rm s}$.

For the parallel configuration, we consider the behavior that the $\Pi/\Pi_{\rm max}$ is higher than $q_0$ increases, with a slower descent until the jet break time, where the polarization plummets. The disallowed set faster decrease would indicate that the polarization at break times must remain high, if we consider the best fit option for the set of Swift-identified bursts is a sufficiently off-axis emission. As such, a polarization $\Pi/\Pi_{\rm max}>80\%$ would be required, at the time of the break, by our model.

\section{Promising GW events in the third observing run (O3) that could generate electromagnetic emission}\label{sec6}

\subsection{Multi-band observations}

During the O3 observing run (from 2019 April 01 to 2020 March 27), the Advanced LIGO and Advanced Virgo GW detectors reported 56 GW events. The run was homogeneously split into two periods called ``O3a'' (from 2019 April 01 to September 30) and ``O3b'' (from 2019 November 01 to 2020 March 27). The candidate GW events in the O3a and O3b runs are reported in Gravitational Wave Transient (GWTC-2) Catalog 2 and (GWTC-3) Catalog 3, respectively. The potential candidates reported that are consistent with a source with $m_2<3M_\odot$ -- where $m_2$ is the mass of the secondary component of the binary merger -- and that could generate electromagnetic emission are GW190425, GW190426\_152155, GW190814 in GWTC-2 \citep{2021PhRvX..11b1053A} and GW191219\_163120, GW200105\_162426, GW200115\_042309, GW200210\_092254 in GWTC-3 \citep{2021arXiv211103606T}. 

\subsection{Light Curves and Polarization}

Figure \ref{GRBs_Swift} second and third rows presents the five promising GW events in the third observing run (O3) which are more likely to generate an electromagnetic counterpart, that is, in whose binary system there is at least one neutron star. Each panel shows the multi-band upper limits and the synchrotron light curves from the off-axis jet decelerating in a constant-density medium with different viewing angles $\theta_{\rm obs}= 6\,$ (solid) and $17\,{\rm deg}$ (dashed). The synchrotron light curves are presented at 1 keV (green), UVOT (orange), R-band (yellow) and 3 GHz (brown). Optical data were retrieved for the follow-up campaign carried out by the DDOTI collaboration \citep{2021MNRAS.507.1401B}. For the chosen values, the values of viewing angle less than $7\, {\rm deg}$ are ruled out in our model for the S190425z (GW190425), S190426C (GW190426\_152155) and S190814bv (GW190814) events which are consistent with the ones reported in \cite{2019ApJ...887L..13D,2020arXiv200201950A, 2019ApJ...884L..55G} using different off-axis models. More observations on duration ranging seconds from the burst trigger to months and years after the merging period are needed to infer tighter constraints.\\

\begin{figure}
{\includegraphics[width=\textwidth]{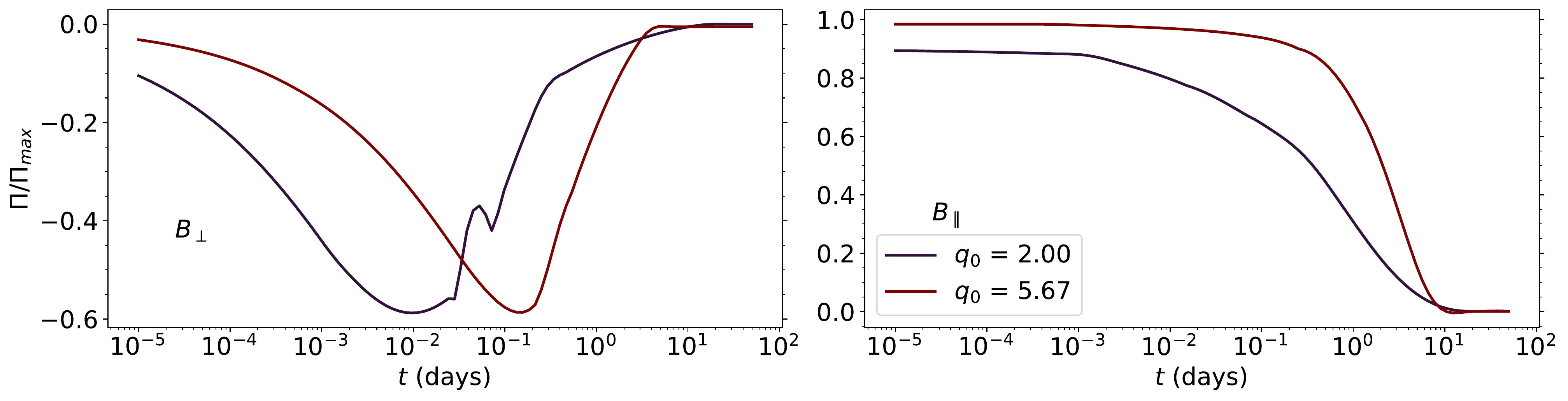}}
\caption{Expected temporal polarization evolution for the set of promising GW events observed in the third run (03) capable of generating electromagnetic emissions. This polarization is calculated requiring the relevant parameters presented in \Cref{tab:Best-Fit-Parameters}}
\label{fig:joined_gw}
\end{figure}

\Cref{fig:joined_gw} shows the expected polarization curves that could be present because of the parameters used to satisfy the upper limits of the GW events.  Similar to the Swift-identified bursts, we will be referring to the two sets of parameters as ``disallowed" (for $\theta_{\rm obs}=6\deg$) and ``allowed" (for $\theta_{\rm obs}=17\deg$).

Similar considerations can be taken as with the Swift-identified bursts; with the intersection happening at $t\approx 3.8\times 10^{-2}\,{\rm s}$ and $\abs{\Pi/\Pi_{\rm max}} \approx 47\%$, we can set the rough upper limit of $< 47\%$ for $t < 3.8\times 10^{-2}\,{\rm s}$, and the requirement of  $\abs{\Pi/\Pi_{\rm max}}> 47\%$ for later times. Furthermore, the narrow $\theta_j$ and large $\theta_{\rm obs}$ constrain the allowed set at $q_0>5$. \cite{2004MNRAS.354...86R} have shown that for a homogeneous sideways expanding jet model, the value of $q_0>5$ threshold leads to a merging of the dual peaks present for $q_0<5$, which is consistent with \Cref{fig:joined_gw} and an extra condition imposed on the polarization for this burst.
Following the same procedure applied for the Swift-identified bursts, a rough limit for the parallel field would be $\Pi/\Pi_{\rm max}>83\%$ at the jet break time.

\section{Conclusions}\label{sec7}

We have introduced a polarization phenomenological model as an extension of the analytical synchrotron afterglow off-axis scenario presented in \cite{2019ApJ...871..200F, 2020ApJ...896...25F}. This synchrotron model can describe the multi-wavelength afterglow observations for both a constant-density and wind-like medium. We have shown the expected temporal evolution of polarization with a dependency on the physical parameters associated with afterglow GRB emission for two configurations of a magnetic field. Regarding our fiducial model, the calculated polarization took into consideration a broad set of parameters constrained within the typical values observed for off-axis GRBs.
We were able to see the differences in possible polarization caused by the two different ambient media and the chosen synchrotron model. We showed that our fiducial model generally agrees with previously found results for a homogeneous sideways expanding jet for the conditions of constant-density medium and adiabatic case with $\xi=1$ \citep{2002ApJ...570L..61G, 2004MNRAS.354...86R}. We have expanded the scenarios for a partially radiative regime and a case where $\xi < 1$. We expect that variation of these parameters present modifications on the temporal evolution of polarization; A partially radiative regime hastens the deceleration of the relativist outflow by the circumburst medium \citep{2000ApJ...532..281B, 2005ApJ...619..968W}, and this has exacerbated the baseline ($\xi=1,\, \epsilon = 0$) profile of polarization -- with peak $\Pi/\Pi_{\rm max}$ increasing further as $q_0$ grows, but second bump decreasing slightly. On the other hand, changing $\xi$ alters the arrival time of the emission \citep{1997ApJ...491L..19W, 2000ApJ...536..195C, 1998ApJ...493L..31P} and our chosen value of $\xi=0.56$ \citep{2000ApJ...536..195C} has caused the polarization behavior regarding $q_0$ to flip, with the magnitude of the peaks now decreasing as $q_0$ increases. Furthermore, we have calculated the same polarization for a wind-like medium to verify the possible differences. For the change in circumburst medium we have found that the polarization evolves slower in time and changes in the magnitude of polarization compared to the constant-density medium, in agreement with \cite{2004A&A...422..121L}.

We have obtained the expected polarization curves for a sample of bursts showing similar off-axis afterglow emissions -- GRB 080503, GRB 140903A, GRB 150101B, GRB 160821B, and GRB 170817A. In particular, we have used the available polarimetric upper limits of GRB 170817A; $\Pi < 12\%$ at 2.8 GHz and $t\approx244$ days \citep{2018ApJ...861L..10C} to rule out our chosen magnetic field configurations of anisotropy factors $b=0$ and $b\rightarrow\infty$. 
Although the remaining bursts have neither detected polarization nor constrained upper limits, from our calculations, we can observe a few patterns that reinforce the similarity between these bursts. For the perpendicular field configuration, GRB 080503 and GRB 140903A showed similar magnitudes of polarization, but somewhat dephased in time. Regarding GRB 150101B, the second peak also has a similar polarization degree to the first one of the previously mentioned bursts; however, the peaks happen much later. GRB 160821B is the most distinct out of these bursts, as the polarization happens considerably faster, with a higher first peak (but not too dissimilar to the previous bursts) and a much lower second peak. This is likely due to the angular properties of the burst, as $q_0$ is closer to unity. GRB 170817A is immersed in a lower external density, with a somewhat more energetic jet seen at wider angles, which in combination causes the peaks to be higher than the other bursts by roughly $10\%$ and happens at later times. The peaks of polarization also roughly coincide with the afterglow flux peak in time \citep[see][for the flux fitting]{2019ApJ...871..200F, 2020ApJ...896...25F}, which is a result that agrees with the literature \citep{1999MNRAS.309L...7G, 2003ApJ...594L..83G, 2004MNRAS.354...86R, 2021MNRAS.507.5340T}. Overall, we could observe the similarities between the bursts' polarization. However, the particularities of each are sufficient to cause observable differences between them.

With the model presented in Section \ref{sec2} and \cite{2020ApJ...896...25F}, we have constrained the possible values of the physical parameters of our system. We take into consideration the upper limits of the four closest sGRBs detected by Swift - GRB 050906, GRB 070810B, GRB 080121 and GRB 100216A - and a set of five GW events that could produce an electromagnetic counterpart - S190425z, S190436c, S190814bv, S200105ae, and S200115j - under the condition they must be narrowly collimated jets and seen sufficiently off-axis. We have obtained two sets of parameters, one allowed by the upper limits and one disallowed, and the projected polarization for these values. We used these two sets to obtain what could be considered as a rough constrain on polarization degree, dependent on the geometry of the magnetic field chosen.

More observations, from seconds after the trigger to months and years, are needed to infer tighter constraints on polarization and adequate fitting of the light curves is necessary to obtain adequate parameter values and break degeneracy between synchrotron models.

\section*{Acknowledgement}

We thank Walas Oliveira, Rodolfo Barniol Duran, Tanmoy Laskar, Paz Beniamini and Bing Zhang for useful discussions. AP acknowledges financial support from CONACyT's doctorate fellowships, NF acknowledges financial support from UNAM-DGAPA-PAPIIT through grant IN106521. RLB acknowledges support from CONACyT postdoctoral fellowships and the support from the DGAPA/UNAM IG100820 and IN105921.

\section*{Data Availability}
The data used for this study was obtained from the respective credited references: upper limits of linear polarization for GRB 170817A \citep[obtained by][]{2018ApJ...861L..10C}; upper limits of GRB 050906, GRB 070810B, GRB 080121, and GRB 100216A \citep[taken from][]{2020MNRAS.492.5011D}; and upper limits for  GW190425, GW190426\_152155, GW190814 in GWTC-2 \citep{2021PhRvX..11b1053A} and GW191219\_163120, GW200105\_162426, GW200115\_042309, GW200210\_092254 in GWTC-3 \citep{2021arXiv211103606T}. Optical upper limits were obtained by \cite{2021MNRAS.507.1401B}.
Other than cited sources, there is no new data generated or analysed in support of this research.

\bibliographystyle{mnras}
\bibliography{references}  
\newpage

\begin{table}
\centering \renewcommand{\arraystretch}{2.2}\addtolength{\tabcolsep}{2.4pt}
\caption{Parameters used to calculate the polarization curves for the fiducial model}\label{tab:table_general}
\begin{tabular}{ c c c c c c c}
\hline
\hline
{\large $\tilde{E}\, (10^{50}\,{\rm erg})$} & {\large ${\rm n}\,\, ({\rm cm^{-3}})$} & {\large ${\rm A}_{\rm w}$\footnote{This value is used for wind-like scenario}} & {\large $\theta_{\rm j}\,\,(\rm deg)$} & {\large $\theta_{\rm obs}\,\,(\rm deg)$} & {\large $\Gamma_0$} \\

\hline \hline
 $1 $ & $10^{-2}$ &$10^{-4}$ & $5.0$ & $[8,15]$ & 100\\
\hline
\end{tabular}
\\
\centering{The range [$8,15$] for $\theta_{\rm obs}$ represents the three chosen values of $\theta_{\rm obs}=[8.0,\, 11.5,\, 15.0$]}
\\
\end{table}

\begin{table}
\centering \renewcommand{\arraystretch}{2.2}\addtolength{\tabcolsep}{2.4pt}
\caption{Posterior distribution for the parameters used to calculate the polarization for our sample of atypical GRBs}\label{tab:par_mcmc}
\begin{tabular}{ l c c c c c c}
\hline
\hline
{\large  Parameters} & {\large $\tilde{E}\, (10^{50}\,{\rm erg})$} & {\large ${\rm n}\,\, (10^{-2}\,{\rm cm^{-3}})$} & {\large $\theta_{\rm j}\,\,(\rm deg)$} & {\large $\theta_{\rm obs}\,\,(\rm deg)$} & {\large $p$} \\

\hline \hline
\\
{\large GRB 080503} & $1.19^{+0.10}_{-0.10}$ & $1.0^{+0.10}_{-0.10}$ & $5.0$ & $12.45^{+0.35}_{-0.35}$\\
{\large GRB 140903A} & $1.50^{+0.49}_{-0.48}\times 10 $ & $1.307^{+0.49}_{-0.48}\times 10^2$ & $5.0$ & $13.4^{+0.5}_{-0.5}$ \\
{\large GRB 150101B} & $1.10^{+0.29}_{-0.30}\times 10$  & $1.0^{+0.29}_{-0.29}$ &  $5.0$ & $21.95^{+0.45}_{-0.45}$ \\
{\large GRB 160821B} & $1.99^{+0.10}_{-0.10}$ & $0.98^{+0.10}_{-0.09}\times 10$  & $5.0$ & $8.0^{+0.3}_{-0.3}$ \\
\hline
\label{tab:pol_grbs}
\end{tabular}
\end{table}

\clearpage

\begin{table}
\centering \renewcommand{\arraystretch}{2.2}\addtolength{\tabcolsep}{2.4pt}
\caption{Values used in the calculation of polarization for the closest sGRBs detected by the Swift satellite and GW events detected by Ligo-Virgo reported on run O3 catalog.}\label{tab:Best-Fit-Parameters}

\begin{tabular}{ccccccc}

\hline
\hline
\multicolumn{1}{l}{\large  Event}  & \multicolumn{1}{l}{\large  $\tilde{E}$ (erg)} & \multicolumn{1}{l}{\large  n (cm$^{-3}$)} & \multicolumn{1}{l}{\large  $\varepsilon_{\rm e}$} & \multicolumn{1}{l}{\large  $\varepsilon_{\rm B}$} & \multicolumn{1}{l}{\large  $\theta_{\rm obs}\,(\deg)$} \\ \hline 
\multirow{2}{*}{GRB 050906} & $10^{50.7}$ & $10^{-1}$ &  $10^{-1}$ &  $10^{-4}$ & 4  \\
                 & $10^{50}$ &  $10^{-1}$ &   $10^{-1}$  &   $10^{-4}$  & 15   \\ 
\multirow{2}{*}{GRB 070810B} & $10^{51.7}$ &  $10^{-1}$ &   $10^{-1}$  &   $10^{-4}$  & 4   \\
                  & $10^{50}$ &  $10^{-1}$ &   $10^{-1}$  &   $10^{-4}$  & 15   \\ 
\multirow{2}{*}{GRB 080121} &  $10^{51.7}$ &  $10^{-1}$ &   $10^{-1}$  &   $10^{-4}$  & 4   \\
                  & $10^{50.7}$  &  $10^{-1}$ &   $10^{-1}$  &   $10^{-4}$  & 15   \\ 
\multirow{2}{*}{GRB 100216A} & $10^{51}$ &  $10^{-1}$ &   $10^{-1}$  &   $10^{-4}$  & 4   \\
                  & $10^{50}$ &  $10^{-1}$ &   $10^{-1}$  &   $10^{-4}$  & 15   \\ \cline{1-6}
\multirow{2}{*}{S190423z} & $10^{51} $  &  1 &  $10^{-1}$    &   $10^{-3}$  &  6   \\
                  & $10^{49}$ & 1  & $10^{-1}$    &   $10^{-3}$  &  17    \\ 
\multirow{2}{*}{S190426c} & $10^{51} $  &  1 &  $10^{-1}$    &   $10^{-3}$  &  6   \\
                  & $10^{49}$ & $10^{-1}$  & $10^{-1}$    &   $10^{-3}$  &  17    \\
\multirow{2}{*}{S190814bv} & $10^{51} $  &  1 &  $10^{-1}$    &   $10^{-2}$  &  6   \\
                  & $10^{48}$ & 1  & $10^{-1}$    &   $10^{-3}$  &  17    \\ 
\multirow{2}{*}{S200105ae} & $10^{51} $  &  1 &  $10^{-1}$    &   $10^{-3}$  &  6   \\
                  & $10^{49}$ & 1  & $10^{-1}$    &   $10^{-3}$  &  17    \\ 
\multirow{2}{*}{S200115j} & $10^{51} $  &  1 &  $10^{-1}$    &   $10^{-2}$  &  6   \\
                  & $10^{49}$ & 1  & $10^{-1}$    &   $10^{-2}$  &  17    \\                   
\hline                  
\end{tabular}
\end{table}

\clearpage


\clearpage

\bsp	
\label{lastpage}
\end{document}